\documentclass[conference]{IEEEtran}

\usepackage{soul}

\IEEEoverridecommandlockouts

\usepackage{cite}
\usepackage{amsmath,amssymb,amsfonts}
\usepackage{amsthm}
\usepackage{algorithm, algpseudocode}
\usepackage{graphicx}
\usepackage{subcaption}
\usepackage{textcomp}
\usepackage{xcolor}
\usepackage[acronym,shortcuts]{glossaries}
\def\BibTeX{{\rm B\kern-.05em{\sc i\kern-.025em b}\kern-.08em
    T\kern-.1667em\lower.7ex\hbox{E}\kern-.125emX}}

\newtheorem{remark}{Remark}

\usepackage[acronym,shortcuts]{glossaries}
\newacronym{AP}{AP}{access point}
\newacronym{CDF}{CDF}{cumulative distribution function}
\newacronym{CSCG}{CSCG}{circularly symmetric complex Gaussian}
\newacronym{DAS}{DAS}{distributed antenna system}
\newacronym{DMRS}{DMRS}{demodulation reference signals}
\newacronym{iid}{i.i.d.}{independent and identically distributed}
\newacronym{LIS}{LIS}{large intelligent surface}
\newacronym{MIMO}{MIMO}{multiple-input multiple-output}
\newacronym{MMSE}{MMSE}{minimum mean-square error}
\newacronym{SE}{SE}{spectral efficiency}
\newacronym{SI}{SI}{side information}
\newacronym{SINR}{SINR}{signal-to-interference-plus-noise ratio}
\newacronym{SNR}{SNR}{signal-to-noise ratio}
\newacronym{UatF}{UatF}{use-and-then-forget}
\newacronym{UE}{UE}{user equipment}
\newacronym{ZF}{ZF}{zero-forcing}
\newacronym{TDD}{TDD}{time-division duplexing}
\newacronym{CSI}{CSI}{channel state information}
\begin{document}

\title{On Optimal Strategies for Joint\\ Reciprocity Calibration in Distributed MIMO
\vspace*{-.4\baselineskip}
}

\author{\IEEEauthorblockN{Kohei Ueda\IEEEauthorrefmark{1}\IEEEauthorrefmark{2}, Anubhab Chowdhury\IEEEauthorrefmark{2}, Koji Ishibashi\IEEEauthorrefmark{1}, and Erik G. Larsson\IEEEauthorrefmark{2}}\\
\IEEEauthorblockA{\IEEEauthorrefmark{1}Advanced Wireless \& Communication Research Center (AWCC), The University of Electro-Communications, \\1-5-1 Chofugaoka, Chofu, Tokyo 182-8585, Japan\\
\IEEEauthorrefmark{2}Department of Electrical Engineering (ISY), Linköping University, 58183 Linköping, Sweden\\
  Emails: \IEEEauthorrefmark{1}\IEEEauthorrefmark{2}k.ueda@awcc.uec.ac.jp,  \IEEEauthorrefmark{2}anuch87@liu.se, \IEEEauthorrefmark{1}koji@ieee.org, and \IEEEauthorrefmark{2}erik.g.larsson@liu.se
\vspace*{-\baselineskip}}
\thanks{The work of K. Ueda was performed while visiting Link\"oping University.}
\thanks{This work was supported in part by the KAW foundation, ELLIIT, the Swedish Research Council (VR), JST CRONOS, Japan Grant Number JPMJCS24N1, and JST BOOST, Japan Grant Number JPMJBS2415.}
}
\maketitle

\begin{abstract}
This paper investigates the impact of reciprocity calibration errors on the downlink spectral efficiency (SE) of multi-user large antenna systems. Specifically, we consider two calibration approaches: (a) \emph{global calibration}, in which all antennas~(can be distributed access-points~(APs)) in the system cooperatively perform calibration, and (b) \emph{local calibration}, wherein only a subset of antennas involved in downlink beamforming performs calibration.
We derive the downlink SE considering the use-and-then-forget bound and side-information bound, and then demonstrate that, when downlink pilots are employed~(in the case of side-information bound), the global calibration outperforms local calibration for arbitrary calibration topologies.
\end{abstract}


\section{Introduction}
In distributed \ac{MIMO} systems, a large number of geographically~(spatially) separated \acp{AP}, with single or multiple antennas, jointly and phase-coherently serve the \acp{UE}, relying on the uplink estimated \ac{CSI}~\cite{Erik_Jorswieck_JSTSP}. For \ac{TDD}-\ac{MIMO} systems, this is commonly known as reciprocity-based coherent beamforming, which in turn is critical to procure the promised gain in the downlink \ac{SE}~\cite{2016_Erik_mMIMObook, 2024_Erik_MassiveSynchrony, Vieira_Erik}.\footnote{Both channel estimation and reciprocity-calibration are essential for coherent, reciprocity-based downlink beamforming. While channel estimation in \ac{TDD} systems is achieved using uplink pilots, over-the-air calibration requires special inter-antenna signaling that disrupts normal TDD frame-structure~\cite{ngo2025breakingtddflowovertheair}.} 
\par
Specifically, the $n$-th antenna is associated with a phase offset $\phi_n$, which sums the phase shifts from both the transmit and receive RF chains and encapsulates the effects of (a) local oscillator phase-drift and (b) phase-lag due to hardware imperfections~\cite{2024_Erik_MassiveSynchrony}.
In a distributed large antenna system, the former can be more critical, especially if the oscillators are not locked to a common
reference. Now, a system is said to be reciprocity calibrated
if $\phi_n$ are known up to a common constant for all $n$. Over-the-air methods for reciprocity calibration
typically use bidirectional measurements between \acp{AP}, essentially
measuring $\phi_n-\phi_{n^\prime}$ for different pairs of $\{n, n^{\prime}\}$; which enables reciprocity-based, joint coherent multiuser
\ac{MIMO} beamforming~\cite{Vieira_Erik}. Detailed analyses of reciprocity calibration methods, specifically for distributed antenna systems, can be found in~\cite{Kunnath, Chang_Chung, Nissel, Balan_Antonios}, and references therein.
\par
This work studies the impact of reciprocity calibration errors on 
\ac{SE} in a multi-user \ac{MIMO} system, extending the framework of \cite{2024_Erik_MassiveSynchrony}, which developed a rigorous framework to analyze the effect of calibration errors on beamforming accuracy.
This paper considers two types of calibration methods: ``global calibration'', in which all antennas in the system are involved in the calibration process, and ``local calibration'', in which only the subset of antennas used for beamforming is involved in the calibration process.
{While global calibration is performed using all \acp{AP} in the entire network, its efficacy is dependent on the network topology.
\cite{2024_Erik_MassiveSynchrony} demonstrated that the variance of calibration errors can scale significantly with the number of antennas involved in the calibration process, depending on the specific calibration topology.}
{Finally, we note that a user-centric approach is often considered optimal in terms of implementing coherent beamforming \cite{Emil_User_Centric}. From a calibration perspective, \cite{2024_Erik_MassiveSynchrony} revealed that global calibration provides theoretically superior coherent beamforming accuracy in forming a spatial null within distributed antenna systems.
However, in multi-\ac{UE} scenarios, calibration errors induce fluctuations not only in the null-forming capability toward other \acp{UE} but also in the effective gain of the desired signal.
Therefore, from the perspective of \ac{SE}, it remains unclear whether global or local calibration is more effective.
}
Our work investigates that framework for the multi-user case and derives achievable \ac{SE} bounds.
Our specific findings are that:
\begin{enumerate}
    \item In the absence of downlink pilots~(also known as \ac{DMRS}), in the large-antenna regime with calibration based on a complete-graph topology, global calibration always achieves higher \ac{SE} than local calibration.
    \item When downlink pilots with reasonably high-\ac{SNR} are employed, the \ac{SE} with global calibration consistently outperforms that of local calibration, for arbitrary calibration topologies.
\end{enumerate}

Through numerical evaluations, we further demonstrate how these analytical findings accurately predict the \ac{SE} behavior in a practical system, and validate the \ac{SE} performance from the perspectives of both ergodic capacity and analytical bounds.
\subsubsection*{Notation}
The operators $(\cdot)^{*}$, $(\cdot)^{\mathrm{T}}$, and $(\cdot)^{\mathrm{H}}$ denote the conjugate, transpose, and Hermitian transpose, respectively.
The expectation, Hadamard product, $\ell_2$-norm are denoted as $\mathbb{E}[\cdot]$, $\odot$, and $\|\cdot\|$, respectively. The identity matrix of size $N$ is denoted as $\mathbf{I}_N$.
$\mathbf{0}_{A\times B}$ and $\mathbf{1}_{A\times B}$ are the all-zero and all-one vector/matrix of size $A\times B$.

\section{System Model}
\label{sec:model}
We consider a \ac{DAS} that consists of $N$ antennas and $K$ single-antenna \acp{UE}.
We model the non-reciprocity phase coefficient of the $n$-th antenna by $\phi_n$, which is defined $\mathrm{mod}$ $2\pi$ \cite{2022_reciprocityNewModel}.
The objective of the reciprocity calibration is to estimate $\hat{\phi}_n$.
\par
Once we obtain the estimate of the reciprocity coefficients, the $N$ antennas cooperatively transmit precoded data to \acp{UE}.
Let $p_n=e^{j\phi_n}$ and $\hat{p}_n=e^{j\hat{\phi}_n}$ be the phase error corresponding to the non-reciprocity, and its estimate, respectively.
We assume that the downlink transmission is performed using only a subset of $N_{\Omega}$ antennas with index $\Omega=\{n_1,\dots,n_{N_\Omega}\} \subset \{1,\dots, N\}$.
Therefore, the beamforming vector for the $k$-th \ac{UE} can be expressed as $\mathbf{f}_{k}=[f_{k1},\dots,f_{k N}]^{\mathrm{T}}\in\mathbb{C}^{N\times 1}$, where $f_{kn} \neq 0$ only for $n \in \Omega$, and $f_{kn} = 0$ otherwise.
We also assume that the beamforming vector is generated with \ac{ZF}.
The channel  between the $k$-th \ac{UE} and antennas is defined as $\mathbf{h}_k = [h_{k1}, \dots,h_{kN}]^{\mathrm{T}}$, where $h_{kn} \neq 0$ only for $n \in \Omega$, and $h_{kn} = 0$ otherwise.
Based on the above definitions, the received signal at the $k$-th \ac{UE} can be expressed as
\begin{align}
    \label{eq:def_receive}
    y_k &=\! \mathbf{h}_k^{\mathrm{T}}\mathrm{diag}(\mathbf{p}^\ast\odot\hat{\mathbf{p}})\mathbf{f}_k x_k \!+\! \sum_{k^\prime \neq k} \mathbf{h}_k^{\mathrm{T}}\mathrm{diag}(\mathbf{p}^\ast\odot\hat{\mathbf{p}})\mathbf{f}_{k^\prime} x_{k^\prime} \!+\! n_k,
\end{align}
where $\mathbf{p} = [p_1,\dots,p_{N}]^{\mathrm{T}}$, and $\hat{\mathbf{p}}=[\hat{p}_1,\dots,\hat{p}_{N}]^{\mathrm{T}}$.
The transmitted symbol of the $k$-th \ac{UE} and noise at the $k$-th \ac{UE} are denoted by $x_k\sim\mathcal{CN}(0, 1)$ and $n_k\sim\mathcal{CN}(0, \sigma^2)$, which are \ac{iid}.
\par
Considering the use of \ac{DMRS}, each \ac{UE} can estimate and utilize the effective channel, which includes the effect of calibration errors and beamforming for signal detection.
Based on the received signal definition in \eqref{eq:def_receive}, the effective channel at the $k$-th \ac{UE} can be written as
\begin{align}
\label{eq:g_def}
    g_k = \mathbf{h}_k^{\mathrm{T}}\mathrm{diag}(\mathbf{p}^\ast\odot\hat{\mathbf{p}})\mathbf{f}_k.
\end{align}
\section{Reciprocity Calibration Framework: Preliminaries}
In this section, we briefly recapitulate the definitions regarding calibration measurements, calibration topologies, and the calibration process proposed in \cite{2024_Erik_MassiveSynchrony}, in order to set the context for the subsequent analysis.
\subsection{Definition of Calibration Measurements}
In the calibration process, $N$ antennas perform bidirectional over-the-air reciprocity calibration measurements one by one.
{Let $L$ define the total number of measurements.
If antennas $n_\ell$ and $n_{\ell}^\prime$ intercommunicate in the $\ell$-th pilot measurement, where $\ell=1,\dots, L$, this measurement can be written as}
\begin{align}
\label{eq:calibration_measurement}
    x_\ell = \phi_{n_{\ell}} - \phi_{n_\ell^\prime} + w_\ell,
\end{align}
where $w_\ell$ is noise with covariance $\mathbf{Q}=\mathrm{Cov}(\mathbf{w})$, $\mathbf{w}=[w_1,\dots,w_L]^{\mathrm{T}}$.
The calibration measurement topology can be represented by an undirected graph $\mathcal{G}$, where each node and edge correspond to an antenna and a measurement, respectively.
Let $\mathbf{B}$ be the $L\times N$ incidence matrix of $\mathcal{G}$, whose $\ell$-th row corresponds to the $\ell$-th edge of $\mathcal{G}$.
By defining $\mathbf{x}=[x_1,\dots,x_L]^{\mathrm{T}}$ and $\boldsymbol{\phi}=[\phi_1,\dots,\phi_N]^{\mathrm{T}}$, the measurement in \eqref{eq:calibration_measurement} is rewritten as
\begin{align}   \label{eq:calibration_measurement_all}   \mathbf{x}=\mathbf{B}\boldsymbol{\phi} + \mathbf{w}.
\end{align}
We consider two types of calibration.
Following~\cite{2024_Erik_MassiveSynchrony}:
\subsubsection{Global Calibration}
All $N$ antennas are involved in the calibration, where the graph is represented by $\mathcal{G}$ and $\mathbf{B}$.
\subsubsection{Local Calibration}
Only a subset $\Omega$ of antennas, which correspond to the antennas used for downlink beamforming, are involved in the calibration.
The corresponding graph $\mathcal{G}_{\Omega}$ is a subgraph of $\mathcal{G}$, with $N$ nodes and $L_{\Omega}$ edges, where $L_{\Omega}$ is the number of edges between nodes in $\Omega$.
Without loss of generality, we assume that the $L_{\Omega}$ last elements of $\mathbf{x}$ contain $\mathbf{x}_{\Omega}$, i.e., $\mathbf{x}_{\Omega}=[\mathbf{0}_{L-L_{\Omega}\times {L-L_{\Omega}}}~\mathbf{I}_{L_{\Omega}}]\mathbf{x}$.
\par
The graph Laplacian including noise covariance corresponding to the graph $\mathcal{G}$ is $\mathbf{L}=\mathbf{B}^{\mathrm{T}}\mathbf{Q}^{-1}\mathbf{B}$,
and can be rewritten as $\mathbf{L}=\mathbf{Z}\boldsymbol{\Lambda}\mathbf{Z}^{\rm T}$,
where $\boldsymbol{\Lambda}$ is an $(N-1)\times(N-1)$ diagonal matrix whose positive diagonal entries are the non-zero eigenvalues of $\mathbf L$, and 
$\mathbf Z \in \mathbb{R}^{N\times(N-1)}$ is a matrix whose columns form an orthonormal basis for the subspace orthogonal to $\mathbf{1}_{N\times 1}$.
Similarly, for the subgraph, $\mathcal{G}_{\Omega}$, we define $\mathbf{L}_{\Omega} = \mathbf{B}_{\Omega}^{\mathrm{T}}\mathbf{Q}_{\Omega}\mathbf{B}_{\Omega}=\mathbf{Z}_{\Omega}\boldsymbol{\Lambda}_{\Omega}\mathbf{Z}_{\Omega}^{\mathrm{T}},$
where $\boldsymbol{\Lambda}_{\Omega}$ and $\mathbf{Z}_{\Omega} \in \mathbb{C}^{N \times (N_{\Omega}-1)}$ are a diagonal matrix and a matrix whose columns form an orthonormal basis for the subspace orthogonal to both $\mathbf{1}_{N\times1}$ and the subspace spanned by the complement set $\Omega^c$. 
We can now formulate the reciprocity calibration problem.
\subsection{Reciprocity Calibration}
\subsubsection{Global Calibration}
Based on the measurement model in \eqref{eq:calibration_measurement_all}, the least-squares problem to estimate the calibration coefficients $\boldsymbol{\phi}$ is $\underset{\boldsymbol{\phi}}{\mathrm{argmin}}~\left\| \mathbf{x} - \mathbf{B}\boldsymbol{\phi} \right\|^2$.
{Since the component proportional to $\mathbf{1}_{N\times 1}$ does not affect the beamforming gain or the resulting \ac{SE}, the solution $\hat{\boldsymbol{\phi}}=[\hat{\phi}_1,\dots, \hat{\phi}_N]^{\mathrm{T}}$~(relevant to \ac{SE}) is uniquely given by} $\hat{\boldsymbol{\phi}}^{\mathrm{G}} = \mathbf{Z}(\mathbf{Z}^{\mathrm{T}}\mathbf{B}^{\mathrm{T}}\mathbf{Q}^{-1}\mathbf{B}\mathbf{Z})^{-1}\mathbf{Z}^{\mathrm{T}}\mathbf{B}^{\mathrm{T}}\mathbf{Q}^{-1}\mathbf{x}$,
with its covariance being $\mathbf{C}_{\hat{\boldsymbol{\phi}}}^{\mathrm{G}} = \mathbf{Z}(\mathbf{Z}^{\mathrm{T}}\mathbf{L}\mathbf{Z})^{-1}\mathbf{Z}^{\mathrm{T}}$.

\subsubsection{Local Calibration}
Similarly, based on the measurement model in \eqref{eq:calibration_measurement}, the least-squares problem for the local calibration, obtained from the subgraph $\mathcal{G}_{\Omega}$, is $\underset{\boldsymbol{\phi}}{\mathrm{argmin}}~\left\| \mathbf{x}_{\Omega} - \mathbf{B}_{\Omega}\boldsymbol{\phi} \right\|^2,$ whose solution being $\hat{\boldsymbol{\phi}}^{\mathrm{L}} = \mathbf{Z}_{\Omega}\left( \mathbf{Z}_{\Omega}^{\mathrm{T}}\mathbf{B}_{\Omega}^{\mathrm{T}}\mathbf{Q}_{\Omega}^{-1}\mathbf{B}_{\Omega}\mathbf{Z}_{\Omega} \right)^{-1}\mathbf{Z}_{\Omega}^{\mathrm{T}}\mathbf{B}_{\Omega}^{\mathrm{T}}\mathbf{Q}^{-1}_{\Omega}\mathbf{x}_{\Omega},$
and its covariance is $\mathbf{C}_{\hat{\boldsymbol{\phi}}}^{\mathrm{L}} = \mathbf{Z}_{\Omega}(\mathbf{Z}_{\Omega}^{\mathrm{T}}\mathbf{L}_{\Omega}\mathbf{Z}_{\Omega})^{-1}\mathbf{Z}_{\Omega}^{\mathrm{T}}$.

Note that the estimation error variance of $\boldsymbol{\phi}$ increases with the number of antennas in some topologies even with the case of global calibration. Thus, it is not prima facie apparent whether the \ac{SE} would be higher with global calibration compared to with local.
These aspects are discussed next.
\section{Spectral Efficiency Analysis}
We now evaluate the effect of the calibration errors arising from the calibration process and compare with the \acp{SE}.
\subsection{Effect of Calibration Errors on Beamforming Gain}
\label{sec:beam_gain}
First, to quantify the effect of the aforementioned calibration errors on the \ac{SE} performance, we summarize the effect of calibration errors on beamforming accuracy analyzed in \cite{2024_Erik_MassiveSynchrony}.
We define the intermediate calibration error term as $\tilde{\phi}_n^S = \hat{\phi}_n^S - \phi_n + \bar{\phi}^S$,
where $\tilde{\phi}_n^S$ and $\bar{\phi}^S$ 
are the zero-mean part of calibration error and the common part of calibration coefficients over all antennas, respectively.
$S\in\{\mathrm{L}, \mathrm{G}\}$ stands for calibration type ``Local'' and ``Global''.
Using the above quantity and system model defined in \eqref{eq:def_receive}, the effective channel gain at the $k$-th \ac{UE} corresponding to the beam for the $k^\prime$-th \ac{UE} can be written as
\begin{align}
    &\gamma^S_{kk^\prime} = \mathbf{h}_k^{\mathrm{T}}\mathrm{diag}(\mathbf{p}^\ast\odot\hat{\mathbf{p}})\mathbf{f}_{k^\prime} = \sum_{n\in\Omega}h_{kn}f_{k^\prime n}e^{-j\phi_n}e^{j\hat{\phi}_n^S} \nonumber\\
    \label{eq:channel_approx}
    &\approx e^{-j\bar{\phi}^S}\sum_{n\in\Omega} h_{kn}f_{k^\prime n}(1 + j\tilde{\phi}_n^S),
\end{align}
where in the second line we use a first-order Taylor expansion.
This expansion is accurate when $\tilde{\phi}_n^S$ is zero-mean and in the limit of weak measurement noise.
\par
Regarding effective gain, assuming effective channel $\mathbf{v}_{kk^\prime}=\mathbf{h}_k\odot\mathbf{f}_{k^\prime}$, the variance of $\gamma^S_{kk^\prime}$ can be written as $\mathrm{Var}(\gamma^S_{kk^\prime}) \approx \mathbf{v}_{kk^\prime}^{\mathrm{H}}\mathbf{C}_{\hat{\phi}}^S\mathbf{v}_{kk^\prime}$, where $\mathbf{1}_{1\times N}\mathbf{v}_{kk^\prime}=0$, which implies that $\mathbf{v}_{kk^\prime}$ forming a spatial null.
It is proved in \cite{2024_Erik_MassiveSynchrony} that the variance of the effective gain $\mathrm{Var}(\gamma^S_{kk^\prime})$ with global calibration is always smaller than that of local calibration, i.e.,
\begin{align}
  \label{eq:variance_beam}
  \mathrm{Var}(\gamma^{\mathrm{L}}_{kk^\prime}) - \mathrm{Var}(\gamma^{\mathrm{G}}_{kk^\prime}) = \mathbf{v}_{kk^\prime}^{\mathrm{T}}\mathbf{C}_{\hat{\phi}}^{\mathrm{L}}\mathbf{v}_{kk^\prime}^\ast - \mathbf{v}_{kk^\prime}^{\mathrm{T}}\mathbf{C}_{\hat{\phi}}^{\mathrm{G}}\mathbf{v}_{kk^\prime}^\ast \ge 0,
\end{align}
which implies that global calibration always gives better beamforming accuracy than that of local calibration, when we place a spatial null.
\par
The analysis presented in \cite{2024_Erik_MassiveSynchrony}, was limited to a single null-forming configuration, i.e., $\mathbf{1}_{1\times N}\mathbf{v}_{kk^\prime}=0$, and focused on the resulting beamforming gain. This work considers a multi-\ac{UE} scenario and analyzes the impact of calibration on the achievable \ac{SE}.
{Moreover, we extend the asymptotic beamforming gain analysis with complete topology to a more general class of vectors $\mathbf{v}_{kk}$ for which $\mathbf{1}_{1\times N}\mathbf{v}_{kk}\neq 0$.}
%
\subsection{SE Performance Without Downlink Pilots}
Using the effective gain definition in \eqref{eq:channel_approx}, the received signal in \eqref{eq:def_receive} with calibration type $S$ can be rewritten as
\begin{align}
    &e^{j\bar{\phi}^S}y_k^S \approx \mathbf{h}_k^{\mathrm{T}}\mathbf{f}_kx_k + j(\mathbf{h}_k\odot\mathbf{f}_k)^{\mathrm{T}}\tilde{\boldsymbol{\phi}}^Sx_k \nonumber\\
    &\hspace{10ex}+ \sum_{k^\prime\neq k}j(\mathbf{h}_k\odot\mathbf{f}_{k^\prime})^{\mathrm{T}}\tilde{\boldsymbol{\phi}}^Sx_{k^\prime} + e^{-j\bar{\phi}^S}n_k\\
    &= \mathbf{h}_k^{\mathrm{T}}\mathbf{f}_kx_k + j\mathbf{v}_{kk}^{\mathrm{T}}\tilde{\boldsymbol{\phi}}^Sx_k + \sum_{k^\prime\neq k}j\mathbf{v}_{kk^\prime}^{\mathrm{T}}\tilde{\boldsymbol{\phi}}^Sx_{k^\prime} + e^{-j\bar{\phi}^S}n_k\nonumber
\end{align}
where $\tilde{\boldsymbol{\phi}}^S=[\tilde{\phi}_1^S, \dots, \tilde{\phi}_{N}^S]^{\mathrm{T}}$ is zero-mean calibration error vector.
Then, in the second line, we use $\mathbf{h}_k^{\mathrm{T}}\mathbf{f}_{k^\prime} = 0, \forall k^\prime \neq k$.
\par
For quantifying the effect of calibration errors on the \ac{SE} performance, assuming ergodicity with respect to all sources of randomness, including the calibration errors
following the \ac{UatF} bound principle \cite[Ch. 3.2]{2016_Erik_mMIMObook} as follows:
\begin{align}
    \label{eq:desire_UatF}
    y_k^S = \mathbb{E}\left[ e^{-j\bar{\phi}}\mathbf{h}_k^{\mathrm{T}}\mathbf{f}_k \right]x_k + z_k^S,
\end{align}
The effective interference and noise term $z_k$ in \eqref{eq:desire_UatF} is
\begin{align}
    \label{eq:noise_UatF}
    z_k^S &= e^{-j\bar{\phi}^S}\left( \mathbf{h}_k^{\mathrm{T}}\mathbf{f}_k - \mathbb{E}\left[ \mathbf{h}_k^{\mathrm{T}}\mathbf{f}_k \right] + j\mathbf{v}_{kk}^{\mathrm{T}}\tilde{\boldsymbol{\phi}}\right)x_k\nonumber\\
    &\hspace{4ex}+ \sum\nolimits_{k^\prime\neq k} e^{-j\bar{\phi}^S+\frac{\pi}{2}}\mathbf{v}_{kk^\prime}^{\mathrm{T}}\tilde{\boldsymbol{\phi}}^Sx_{k^\prime} + n_k,
\end{align}
With these, a lower bound on the (ergodic) \ac{SE} of the $k$-th \ac{UE} is $R_{\mathrm{U}, k}^S = \log(1+\mathrm{SINR}_{\mathrm{U},k}^S)$, where the 
effective \ac{SINR} is computed as
\begin{align}
    \label{eq:def_SINR}
    &\mathrm{SINR}_{\mathrm{U},k}^S = \frac{\mathrm{GI}_k}{\mathrm{BI}_k + \mathrm{CI}_k^S + \mathrm{MI}_k^S + \sigma^2},
\end{align}
where $\mathrm{GI}_k = \left| \mathbb{E}\left[ \mathbf{h}_k^{\mathrm{T}}\mathbf{f}_k \right] \right|^2$, $
    \mathrm{BI}_k = \mathrm{Var}(\mathbf{h}_k^{\mathrm{T}}\mathbf{f}_k)$, $
    \mathrm{CI}_k^S = \mathbb{E}[ | \mathbf{v}_{kk}^{\mathrm{T}}\tilde{\boldsymbol{\phi}}^S |^2 ]$, and $
    \mathrm{MI}_k^S = \sum_{k^\prime\neq k}\mathbb{E}[|\mathbf{v}_{kk^\prime}^{\mathrm{T}}\tilde{\boldsymbol{\phi}}^S|^2].$
Here, the expectations of the above variables are taken over the channel $\mathbf{h}_k$ and measurement noise on the calibration measurements in \eqref{eq:calibration_measurement}.
Based on the independence between $\mathbf{v}_k$ and $\tilde{\boldsymbol{\phi}}^S$, the expectation of the denominator can be rewritten as
\begin{align}
    \label{eq:expectation}
    \hspace*{-1mm}
    \mathbb{E}\left[ |\mathbf{v}_{kn}^{\mathrm{T}}\tilde{\boldsymbol{\phi}}^S|^2 \right] \hspace*{-1mm}=\hspace*{-1mm} \mathbb{E}\left[ \mathbf{v}_{kn}^{\mathrm{T}}\tilde{\boldsymbol{\phi}}^S(\tilde{\boldsymbol{\phi}}^S)^{\mathrm{H}}\mathbf{v}_{kn}^\ast \right]\hspace*{-1mm} =\hspace*{-1mm} \mathbb{E}\left[ \mathbf{v}_{kn}^{\mathrm{T}}\mathbf{C}_{\tilde{\phi}}^S\mathbf{v}_{kn}^\ast \right]\hspace*{-1mm}.
\end{align}
According to \eqref{eq:def_SINR} and the definition of $\mathbf{v}_{kn}$, 
 $\mathbf{1}_{1\times N_{\Omega}}\mathbf{v}_{kn}=0$ holds only for $\mathrm{MI}_k^S$, 
where $\mathbf{v}_{kn}$ is associated with a \ac{ZF} beamforming vector. 
 Therefore, from \eqref{eq:variance_beam}, we get
\begin{align}
    \label{eq:MI_relation}
    \mathrm{MI}_k^{\mathrm{L}}\ge \mathrm{MI}_k^{\mathrm{G}}.
\end{align}

However, for $\mathrm{CI}_k^S$, $\mathbf{v}_{kk}$ is given by the element-wise product of the $k$-th \ac{UE}'s channel and the beamforming, 
for which the ZF orthogonality does not hold.\footnote{We speculate that this can be an artifact of the \ac{UatF} bounding technique, and whether a more general result can be derived is yet to be investigated.} 
Hence, no definite ordering between $\mathrm{CI}_k^L$ and $\mathrm{CI}_k^G$ can be established in general in the finite-dimensional setting. However, when $N\rightarrow \infty$, $N_{\Omega}\rightarrow \infty$ while $N_\Omega<N$, with a complete-graph calibration topology, it holds that $\mathrm{CI}_k^{\mathrm{L}} - \mathrm{CI}_k^{\mathrm{G}} > 0$ for arbitrary beamforming vector $\mathbf{f}_k$ and combined effective channel $\mathbf{v}_{kk}$. The detailed derivation is provided in Appendix~\ref{app:large_diff}.

Therefore, together with the fact that $\mathrm{MI}_k^{\mathrm{L}}-\mathrm{MI}_k^{\mathrm{G}}\ge 0$, this implies that the \ac{UatF} lower bound on the \ac{SINR} with global calibration is always better than that with local calibration $\mathrm{SINR}_{\mathrm{U}, k}^{\mathrm{G}} - \mathrm{SINR}_{\mathrm{U}, k}^{\mathrm{L}} > 0$, in large antenna settings.
\subsection{SE Performance With Downlink Pilots~(i.e., \ac{DMRS})}
Contrary to the previous analyses, we now evaluate the \ac{SE} with the downlink \ac{DMRS}, where the \ac{UE} can utilize an estimate of $g_k$, denoted by $\hat{g}_k$.
Using this estimate $\hat{g}_k$, the received signal can be rewritten as
\begin{align}
    &y_k = g_kx_k + \sum_{k^\prime \neq k} \mathbf{h}_k^{\mathrm{T}}\mathrm{diag}(\mathbf{p}^\ast\odot\hat{\mathbf{p}})\mathbf{f}_{k^\prime} x_{k^\prime} + n_k\nonumber\\
    \label{eq:receive_g}
    &= \hat{g}_kx_k + e_kx_k + \sum_{k^\prime \neq k} \mathbf{h}_k^{\mathrm{T}}\mathrm{diag}(\mathbf{p}^\ast\odot\hat{\mathbf{p}})\mathbf{f}_{k^\prime} x_{k^\prime} + n_k.
\end{align}
Let $e_k=g_k-\hat{g}_k$
be the estimation error corresponding to the equivalent scalar channel coefficient of the $k$-th \ac{UE}.
In this paper, we assume that the \ac{DMRS} are mutually orthogonal and let $\hat{g}_k\in\mathbb{C}$ denote the \ac{MMSE} estimate of $g_k$.
Based on \eqref{eq:receive_g}, the side-information bound~\cite[Ch. 2.3]{2016_Erik_mMIMObook} (treating   $\hat{g}_k$ as the side information) 
gives the following lower bound on (ergodic) capacity:
\begin{align}
  R_{\mathrm{SI}, k}^S &= \mathbb{E}\left[ \log_2\left( 1 + \mathrm{SINR}_{\mathrm{SI}, k}^S \right) \right],\label{eq: R_SI}
\end{align}
where
\begin{align}
    \label{eq:SINR_SI}
    \hspace*{-5mm}\mathrm{SINR}_{\mathrm{SI}, k}^S \hspace*{-1mm}= \hspace*{-1mm}\frac{|\hat{g}_k|^2}{\mathbb{E}\left[ |e_k|^2\middle| \hat{g}_k \right] + \sum_{k^\prime\neq k}\mathbb{E}\left[|\mathbf{v}_{kk^\prime}^{\mathrm{T}}\tilde{\boldsymbol{\phi}}^S|^2\Big|\hat{g}_k\right] \hspace*{-1mm}+ \sigma^2}.\hspace*{-4.5mm}
\end{align}
The outer expectation in~\eqref{eq: R_SI} is taken over $\hat{g}_k$, while the inner conditional expectation $\mathbb{E}\left[ |e_k|^2\middle| \hat{g}_k \right]$ is taken over effective channel $g_k$ and \ac{DMRS} measurement noise.
{The expectation of the term $\mathbb{E}[|\mathbf{v}_{kk^\prime}^{\mathrm{T}}\tilde{\boldsymbol{\phi}}^S|^2|\hat{g}_k]$ is taken over the channel and measurement noise of calibration process, conditioned on $\hat{g}_k$.
\par
To facilitate an analytical comparison between the global and local calibrations, we make a few
approximations. We assume that the equivalent channel $g_k$ follows a \ac{CSCG} distribution, and that $e_k$ and $\hat{g}_k$ are statistically independent, implying $\mathbb{E}\left[ |e_k|^2\middle| \hat{g}_k \right]=\mathbb{E}\left[ |e_k|^2\right]$.
In the conditional expectation of the second term of the denominator in \eqref{eq:SINR_SI}, $\mathbf{v}_{kk^\prime}$ and $\hat{g}_k$ are coupled through the common channel realization $\mathbf{h}_k$ in the \ac{ZF} weights.
However, under an independence approximation, we can write $\mathbb{E}[|\mathbf{v}_{kk^\prime}^{\mathrm{T}}\tilde{\boldsymbol{\phi}}^S|^2|\hat{g}_k] \approx \mathbb{E}[|\mathbf{v}_{kk^\prime}^{\mathrm{T}}\tilde{\boldsymbol{\phi}}^S|^2]$. Note that this approximation is accurate in the large-antenna regime \cite{2013_Hoydis_mMIMO, 2018_Adve_nonreciprocity}.
With these assumptions, the \ac{SINR} in \eqref{eq:SINR_SI} can be rewritten as $\mathrm{SINR}_{\mathrm{SI}, k}^S \approx \dfrac{|\hat{g}_k|^2}{\mathbb{E}\left[ |e_k|^2 \right] + \mathrm{MI}_k^S + \sigma^2}.$
Then, based on the definition of $g_k$ in \eqref{eq:g_def}, the expectation $\mathbb{E}\left[ |e_k|^2\right]$ can be written as follows $\mathbb{E}\left[ |e_k|^2\right]= \frac{\mathrm{Var}(g_k)\sigma^2_{\mathrm{p}}}{\rho_{\mathrm{p}}\mathrm{Var}(g_k) + \sigma_{\mathrm{p}}^2}\approx \frac{\sigma_{\rm p}^2}{\rho_{\rm p}},$
where we approximate the expression by assuming a high-\ac{SNR} regime for the \ac{DMRS} transmission.
Therefore, since $\mathrm{MI}_k^{\mathrm{L}}\ge \mathrm{MI}_k^{\mathrm{G}}$ as shown in \eqref{eq:MI_relation}, the resulting \ac{SINR} with \emph{global calibration is always superiror than that of local calibration.}
}

\begin{remark}
The above analysis reveals that, in the absence of downlink pilots, the resulting bound tends to overestimate the self-interference, such that the superiority of global calibration over local calibration holds only in limited scenarios. In contrast, when downlink pilots are employed, the derived approximated bound guarantees that global calibration outperforms local calibration in the high-SNR regime, regardless of the calibration topology. 
\end{remark}
\section{Numerical Results}
In this section, we provide the numerical results for different calibration topologies.
The considering system consists of $N=64$ antennas, $K=4$ \acp{UE}, and $N_{\Omega}=16$ antennas for beamforming.
Here, the $N_{\Omega}$ \acp{AP} are deployed at the points of a mesh grid within a $500\times500$~meter squared communication area.
For the numerical evaluation, we define the calibration SNR as $\mathrm{SNR}_{\mathrm{cal}} = 1/\sigma_w^2$ and the downlink pilot SNR as $\mathrm{SNR}_{\mathrm{p}} = \rho_{\mathrm{p}}/\sigma_{\mathrm{p}}^2 = 20$~dB. Moreover, the noise covariance matrix is set to $\mathbf{Q} = \sigma_w^2 \mathbf{I}$.
We assume that the channel $\mathbf{h}_k$ follows the \ac{CSCG} distribution with covariance matrix $\mathbf{R}_k$.
The covariance matrix is defined as $\mathbf{R}_k=\mathrm{diag}(\beta_{k,1},\dots,\beta_{k,N})$, where $\beta_{kn}=(d_0/d_{kn})^{-2}$, $d_0=10$~meter, and $d_{kn}$ are large-scale fading coefficient, reference distance, and distance between the $k$-th \ac{UE} and the $n$-th \ac{AP}.
Nevertheless, we underscore that the above analysis remains valid for an arbitrary channel covariance matrix.
The expectations for each bound and the ergodic capacity are numerically evaluated by averaging over $10^3$ Monte Carlo realizations.
\begin{figure*}[t]
    \centering
    \begin{subfigure}{0.32\linewidth}
        \centering
        \includegraphics[width=\linewidth]{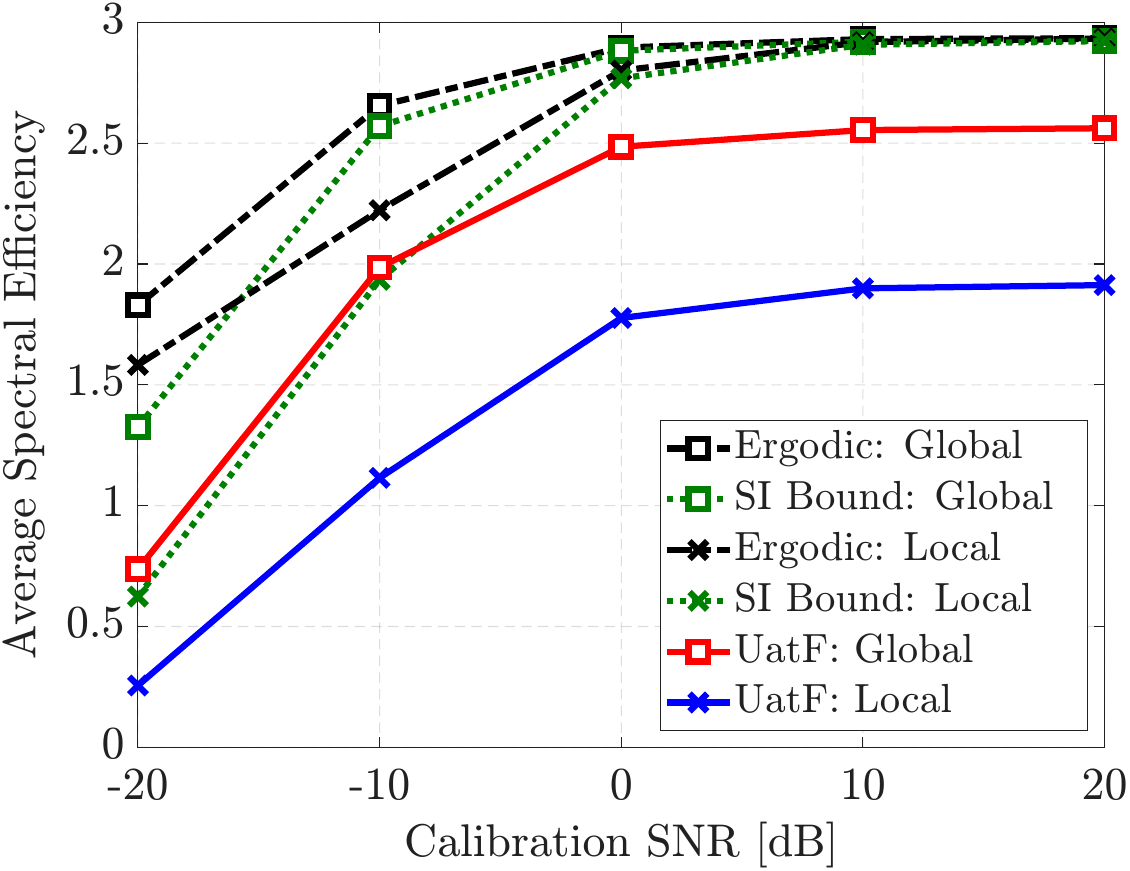}
        \caption{Average SE: Complete graph topology.}
        \label{fig:SE_complete}
    \end{subfigure}
    \hfill
    \begin{subfigure}{0.32\linewidth}
        \centering
        \includegraphics[width=\linewidth]{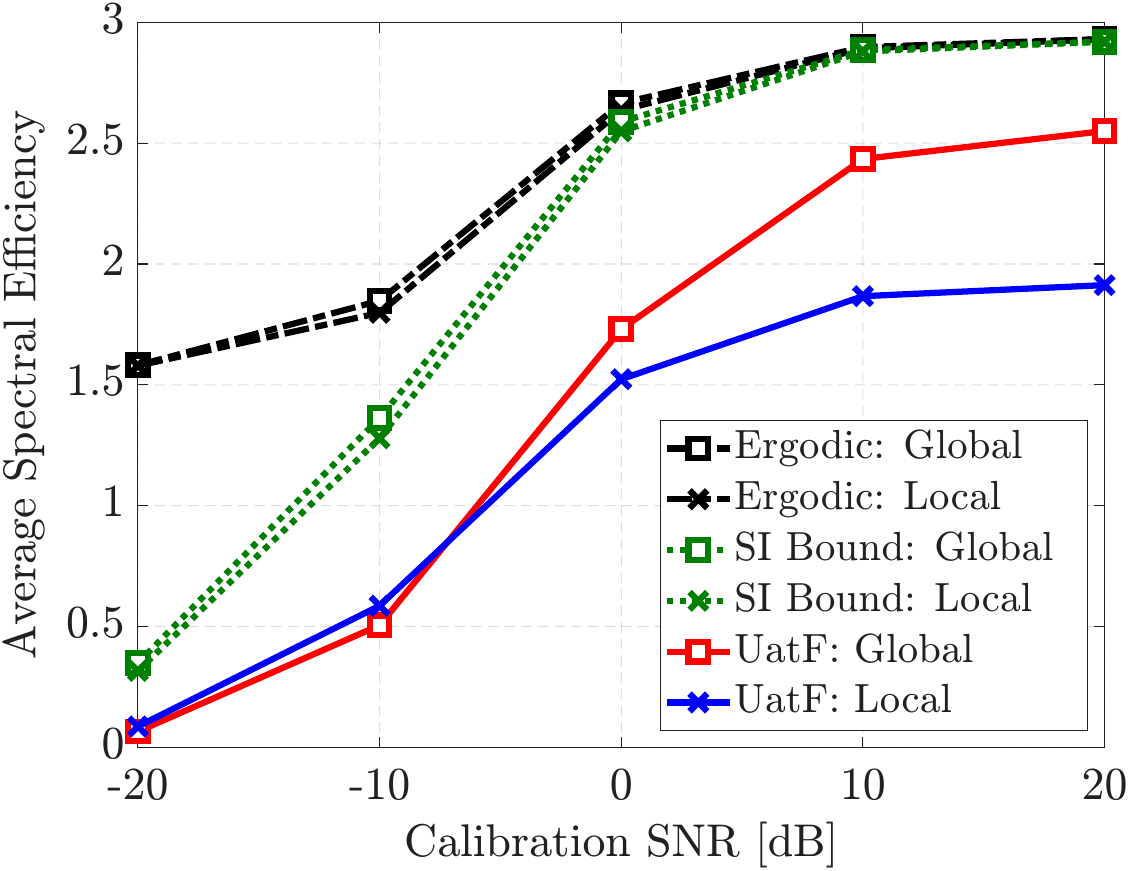}
        \caption{Average SE: LIS topology.}
        \label{fig:SE_LIS}
    \end{subfigure}
    \hfill
    \begin{subfigure}{0.32\linewidth}
        \centering
        \includegraphics[width=\linewidth]{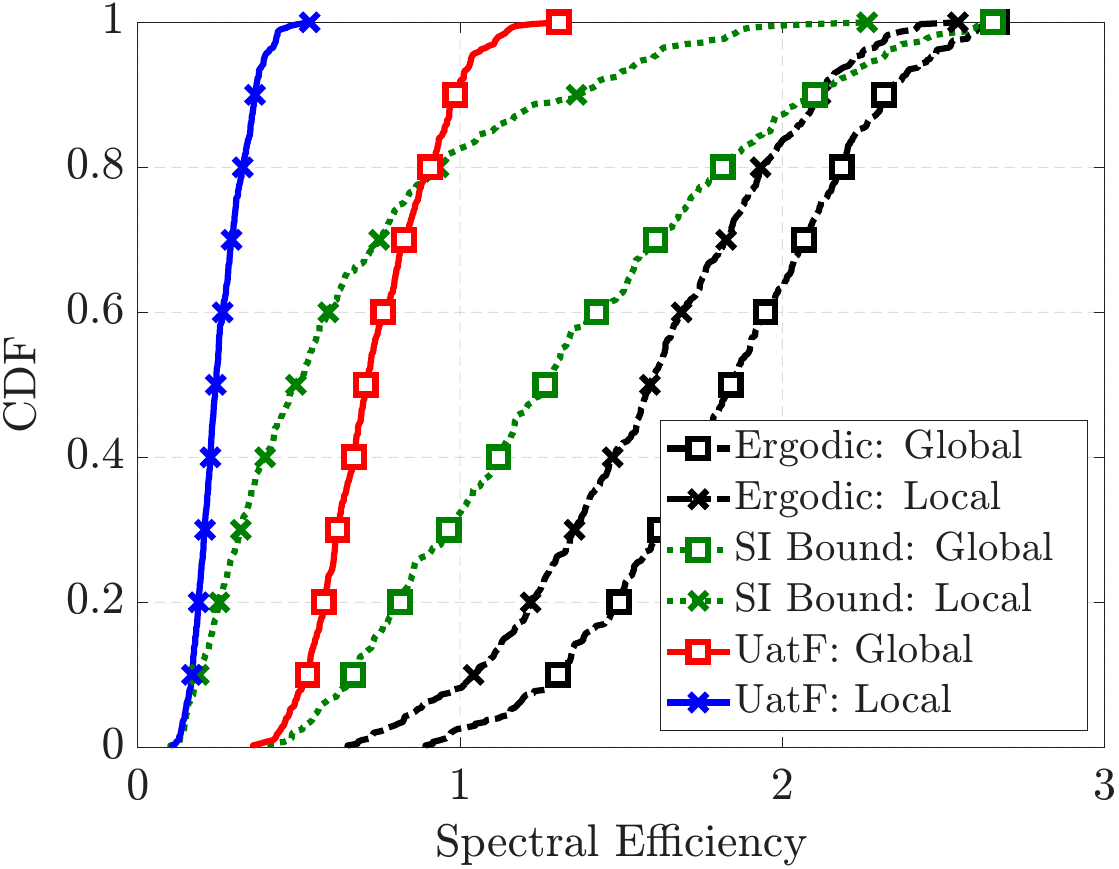}
        \caption{CDF of \ac{SE}: Complete graph topology.}
        \label{fig:SE_CDF}
    \end{subfigure}
    \caption{Average \ac{SE} as a function of calibration \ac{SNR} with different calibration topologies and CDF of \ac{SE} performance with complete graph topology when calibration SNR is $-10$~dB.}
    \label{fig:heatmap}
\end{figure*}
\par

Fig.~\ref{fig:SE_complete} demonstrates the variation in average \ac{SE} over different \acp{UE} as a function of the calibration \ac{SNR} for the complete graph topology. We observe that global calibration consistently achieves higher \ac{SE} than local calibration, both in terms of the analytical bounds and the empirical simulation results. In the high calibration \ac{SNR} regime, both global and local calibration attain high phase calibration accuracy, and thus the procured \acp{SE} are nearly the same for both cases.
\par
Next, Fig.~\ref{fig:SE_LIS} illustrates the \ac{SE} as a function of the calibration \ac{SNR} for the \ac{LIS} topology. Each antenna performs calibration measurements with its closest neighbors in the north–south, east–west, southeast–northwest, and southwest–northeast directions. Here,
we observe that the performance gap between local and global calibration is smaller than that in the complete graph case. This is because, in the complete graph topology, the number of antennas involved in calibration increases linearly as the number of antennas $N$, whereas in the \ac{LIS} topology, this number remains fixed.
Moreover, in the low-SNR regime, the \ac{SE} evaluated based on the \ac{UatF} bound for global calibration becomes lower than that for local calibration. This behavior indicates that, due to the impact of self-interference ``$\mathrm{CI}_k^S$" caused by calibration errors~(which is more an artifact of the bounding technique, rather than the calibration process), the ordering of the \acp{SE} performance derived from the \ac{UatF} bound is not uniquely determined in the arbitrary topology. 
\par
Finally, Fig.~\ref{fig:SE_CDF} presents the \ac{CDF} of the \ac{SE} (considering multiple user locations) for the complete graph topology with a fixed calibration SNR of $-10$~dB, wherein we observe that the ergodic \ac{SE} performance is accurately captured by the side-information bound for both calibration methods.

\section{Conclusion}
Previous work~\cite{2024_Erik_MassiveSynchrony} showed that with joint reciprocity calibration based on bidirectional phase measurements between access points, the phase estimation errors grow, in some cases without bound, the more access points are involved in the calibration process. Despite this, for the purpose of placing spatial nulls,~\cite{2024_Erik_MassiveSynchrony} established through rigorous analysis the optimality of computing a single set of calibration coefficients for the entire network, and using these coefficients in the beamforming. In this paper, we have established that in multi-user MIMO, the same conclusion holds for an achievable bound on the spectral efficiency. 

Our analysis was limited to the cases of certain topologies and two capacity-bounding techniques, a limitation that may be relaxed in future work.
Also, the development of analyses for finite-dimensional cases, followed by robust downlink precoder design can be of interest.

\appendix
\subsection{Proof of $\mathrm{CI}_k^{\mathrm{L}} - \mathrm{CI}_k^{\mathrm{G}} > 0$ in the large antenna regime:}
\label{app:large_diff}
In this proof, we do not impose any specific structural constraints on the equivalent channel vector $\mathbf{v}_{kk}$, except that its support is restricted to the index set $\Omega$ and that it has bounded energy, i.e., $\|\mathbf{v}_{kk}\|^2 = \mathcal{O}(N_{\Omega})$.
{In addition, we assume that the sum of the $\mathbf{v}_{kk}$ converges to arbitrary constant when $N_{\Omega}\rightarrow \infty$, i.e., $|\mathbf{1}_{1\times N}\mathbf{v}_{kk}|/N_{\Omega}\overset{\mathrm{a.s.}}{\longrightarrow}\mathtt{C}$, where $\mathtt{C}$ is a constant, can be mean of the underlying random variables.}
These conditions are typically satisfied in practical beamforming schemes.
In the complete graph topology, the covariance matrices $\mathbf{C}_{\hat{\phi}}^{\mathrm{L}}$ and $\mathbf{C}_{\hat{\phi}}^{\mathrm{G}}$ can be written as follows:
\begin{align}
    \mathbf{C}_{\hat{\phi}}^{\mathrm{G}} &= \frac{1}{N}(\mathbf{I}_N - \frac{1}{N}\mathbf{1}_{N\times N}),\quad\mathbf{C}_{\hat{\phi}}^{\mathrm{L}} = \begin{bmatrix}
        \mathbf{C}_{\mathrm{L}, 1} & \mathbf{0}\\
        \mathbf{0} & \mathbf{0}
    \end{bmatrix},
\end{align}
where $\mathbf{C}_{\mathrm{L}, 1} = \frac{1}{N_{\Omega}}(\mathbf{I}_{N_{\Omega}} - \frac{1}{N_{\Omega}}\mathbf{1}_{N_{\Omega}\times N_{\Omega}})$.
In the large-antenna regime, the second term in $\mathbf{C}_{\hat{\phi}}^{\mathrm{G}}$ scales as $\mathcal{O}(1/N^2)$, whereas the first term scales as $\mathcal{O}(1/N)$.
Hence, as $N\rightarrow \infty$, the second term vanishes faster and $\mathbf{C}_{\hat{\phi}}^{\mathrm{G}}$ can be approximated as $\mathbf{C}_{\hat{\phi}}^{\mathrm{G}}\approx \frac{1}{N}\mathbf{I}_N$.
Similarly, as $N_{\Omega}\rightarrow \infty$, $\mathbf{C}_{\mathrm{L}, 1}$ can be approximated as $\mathbf{C}_{\mathrm{L}, 1} \approx \frac{1}{N_{\Omega}}\mathbf{I}_{N_{\Omega}}$.
Based on the above approximations, we have
\label{eq:CI_inside}
\begin{align}
    &\mathbf{v}_{kk}^{\mathrm{T}}\mathbf{C}_{\hat{\phi}}^{\mathrm{G}}\mathbf{v}_{kk}^\ast \approx \frac{\|\mathbf{v}_{kk}\|^2}{N},\quad \mathbf{v}_{kk}^{\mathrm{T}}\mathbf{C}_{\hat{\phi}}^{\mathrm{L}}\mathbf{v}_{kk}^\ast \approx \frac{\|\mathbf{v}_{kk}\|^2}{N_{\Omega}}.
 \end{align}
Then, the difference between the local and global calibration becomes
\begin{align}
    \hspace*{-.4em}\Delta &= \mathbf{v}_{kk}^{\mathrm{T}}\mathbf{C}_{\hat{\phi}}^{\mathrm{L}}\mathbf{v}_{kk}^\ast - \mathbf{v}_{kk}^{\mathrm{T}}\mathbf{C}_{\hat{\phi}}^{\mathrm{G}}\mathbf{v}_{kk}^\ast = \|\mathbf{v}_{kk}\|^2\left( \frac{1}{N_{\Omega}} - \frac{1}{N} \right),
\end{align}
Since $N_{\Omega} < N$, in a large-antenna regime with a complete calibration topology,
$\mathrm{CI}_k^{\mathrm{L}} - \mathrm{CI}_k^{\mathrm{G}} > 0$ for arbitrary beamforming vector $\mathbf{f}_k$ and combined effective channel $\mathbf{v}_{kk}$.\qed

\bibliographystyle{IEEEtran}
\bibliography{listofpublications}

@ARTICLE{2024_Erik_MassiveSynchrony,
  author={Larsson, Erik G.},
  journal={IEEE Trans. Signal Process.}, 
  title={Massive Synchrony in Distributed Antenna Systems}, 
  year={2024},
  volume={72},
  number={},
  pages={855-866},
  month={Jan.}
}

@article{Emil_User_Centric,
author = {Demir, \"{O}zlem Tugfe and Bj\"{o}rnson, Emil and Sanguinetti, Luca},
title = {Foundations of User-Centric Cell-Free Massive {MIMO}},
year = {2021},
issue_date = {Jan 2021},
publisher = {Now Publishers Inc.},
address = {Hanover, MA, USA},
volume = {14},
number = {3–4},
issn = {1932-8346},
url = {https://doi.org/10.1561/2000000109},
doi = {10.1561/2000000109},
journal = {Found. Trends Signal Process.},
month = jan,
pages = {162–472},
numpages = {315}
}

@ARTICLE{2022_reciprocityNewModel,
  author={Nissel, Ronald},
  journal={IEEE Commun. Lett.}, 
  title={Correctly Modeling {TX and RX} Chain in (Distributed) Massive {MIMO}—New Fundamental Insights on Coherency}, 
  year={2022},
  volume={26},
  number={10},
  pages={2465-2469},
  month={Oct.}
}

@book{2016_Erik_mMIMObook,
  place={Cambridge},
  title={Fundamentals of Massive MIMO},
  publisher={Cambridge University Press},
  author={Marzetta, Thomas L. and Larsson, Erik G. and Yang, Hong and Ngo, Hien Quoc},
  year={2016}
}

@ARTICLE{Erik_Jorswieck_JSTSP,
  author={Xu, Yanqing and others},
  journal={IEEE J. Sel. Top. Signal Process.}, 
  title={Distributed Signal Processing for Extremely Large-Scale Antenna Array Systems: State-of-the-Art and Future Directions}, 
  year={2025},
  volume={19},
  number={2},
  pages={304-330},
  doi={10.1109/JSTSP.2025.3541386},
  ISSN={1941-0484},
  month={Mar.},}

@ARTICLE{2013_Hoydis_mMIMO,
  author={Hoydis, Jakob and ten Brink, Stephan and Debbah, Merouane},
  journal={IEEE J. Sel. Areas Commun.}, 
  title={Massive {MIMO} in the {UL/DL} of Cellular Networks: {How} Many Antennas Do We Need?}, 
  year={2013},
  volume={31},
  number={2},
  pages={160-171},
  month={Feb.}
}

@ARTICLE{2018_Adve_nonreciprocity,
  author={Minasian, Arin and others},
  journal={IEEE Trans. Commun.}, 
  title={Distributed Massive {MIMO} Systems With Non-Reciprocal Channels: Impacts and Robust Beamforming}, 
  year={2018},
  volume={66},
  number={11},
  pages={5261-5277},
  doi={10.1109/TCOMM.2018.2859937},
  ISSN={1558-0857},
  month={Nov.}
}

@INPROCEEDINGS{Vieira_Erik,
  author={Vieira, Joao and Larsson, Erik G.},
  booktitle={Proc. IEEE 32nd
Annu. Int. Symp. Personal, Indoor Mobile Radio Commun. (PIMRC)}, 
  title={Reciprocity calibration of Distributed Massive {MIMO} Access Points for Coherent Operation}, 
  year={2021},
  volume={},
  number={},
  pages={783-787},
  doi={10.1109/PIMRC50174.2021.9569495},
  ISSN={2166-9589},
  month={Sep.},}

@ARTICLE{Kunnath,
  author={Kunnath Ganesan, Unnikrishnan and others},
  journal={IEEE Trans. Wireless Commun.}, 
  title={BeamSync: Over-the-Air Synchronization for Distributed Massive {MIMO} Systems}, 
  year={2024},
  volume={23},
  number={7},
  pages={6824-6837},
  doi={10.1109/TWC.2023.3335089},
  ISSN={1558-2248},
  month={Jul.},}

@ARTICLE{Chang_Chung,
  author={Kim, Nam-I and others},
  journal={IEEE Commun. Lett.}, 
  title={A Gradual Method for Channel Non-Reciprocity Calibration in Cell-Free Massive {MIMO}}, 
  year={2022},
  volume={26},
  number={11},
  pages={2779-2783},
  doi={10.1109/LCOMM.2022.3196928},
  ISSN={1558-2558},
  month={Nov.},}

@ARTICLE{Nissel,
  author={Nissel, Ronald},
  journal={IEEE Commun. Lett.}, 
  title={Correctly Modeling {TX} and {RX} Chain in (Distributed) Massive {MIMO}—New Fundamental Insights on Coherency}, 
  year={2022},
  volume={26},
  number={10},
  pages={2465-2469},
  doi={10.1109/LCOMM.2022.3189968},
  ISSN={1558-2558},
  month={Oct.},}

@ARTICLE{Balan_Antonios,
  author={Balan, Horia Vlad and others},
  journal={IEEE/ACM Trans. Netw}, 
  title={AirSync: Enabling Distributed Multiuser {MIMO} With Full Spatial Multiplexing}, 
  year={2013},
  volume={21},
  number={6},
  pages={1681-1695},
  doi={10.1109/TNET.2012.2230449},
  ISSN={1558-2566},
  month={Dec.},}

@INPROCEEDINGS{ngo2025breakingtddflowovertheair,
  author={Ngo, Khac-Hoang and Larsson, Erik G.},
  booktitle={Proc. IEEE Global Commun. Conf.}, 
  title={Breaking the {TDD} Flow for Over-the-Air Phase Synchronization in Distributed Antenna Systems}, 
  year={2025},
  month={Dec.},
  volume={},
  number={},
  pages={1645-1650}
}
\end{document}